\documentclass[aps,prd,prev,preprintnumbers,floatset,nofootinbib,usenatbib,twocolumn,a4]{revtex4}

\usepackage{blindtext}
\usepackage[a4paper, total={7in, 9in}]{geometry}
\usepackage{graphicx}
\usepackage{dcolumn}
\usepackage{bm}
\usepackage{epsfig}
\usepackage{amsfonts}
\usepackage{amsmath}
\usepackage{amssymb}
\usepackage[usenames]{color}
\usepackage[dvipsnames]{xcolor}
\usepackage[unicode, colorlinks=true, linkcolor=linkcolor, citecolor=linkcolor, filecolor=linkcolor,urlcolor=linkcolor, pdfusetitle]{hyperref}
\usepackage{listings}

\hypersetup{colorlinks,citecolor=blue,linkcolor=blue,urlcolor=blue}
\hypersetup{final=true}

\definecolor{lgreen}{HTML}{079d91}

\begin{document}

\title{Measuring the Hubble Constant with cosmic chronometers: a machine learning approach}

\author{Carlos Bengaly$^1$}
\email{carlosbengaly@on.br}

\author{Maria Aldinez Dantas$^2$}
\email{aldinezdantas@uern.br}

\author{Luciano Casarini$^3$}
\email{lcasarini@academico.ufs.br}

\author{Jailson Alcaniz$^1$}
\email{alcaniz@on.br}

\affiliation{$^1$Observatório Nacional, 20921-400, Rio de Janeiro, RJ, Brazil}
\affiliation{$^2$Departamento de F\'isica, Universidade do Estado do Rio Grande do Norte, 59610-210, Mossor\'o - RN, Brazil}
\affiliation{$^3$Departamento de F\'isica, Universidade Federal de Sergipe, 49000, S\~ao Crist\'ov\~ao - SE, Brazil}

\date{\today}

\begin{abstract}
{Local measurements of the Hubble constant ($H_0$) based on Cepheids e Type Ia supernova differ by $\approx 5 \sigma$ from the estimated value of $H_0$ from Planck CMB observations under $\Lambda$CDM assumptions. In order to better understand this $H_0$ tension, the comparison of different methods of analysis will be fundamental to interpret the data sets provided by the next generation of surveys.} In this paper, we deploy machine learning algorithms to measure the $H_0$ through a regression analysis on synthetic data of the expansion rate  assuming {different values of redshift and different levels of uncertainty}. We {compare the performance of different regression algorithms as Extra-Trees, Artificial Neural Network, Gradient Boosting, Support Vector Machines}, and we find that the Support Vector Machine {exhibits the best performance in terms of bias-variance tradeoff in most cases}, {showing itself a competitive cross-check to non-supervised regression methods such as Gaussian Processes.}

\end{abstract}


\maketitle


\section{Introduction}\label{sec:intro}

The standard model of Cosmology consists of a flat, homogeneous and isotropic universe whose energy content is dominated by a cosmological constant ($\Lambda$) and cold dark matter ($\Lambda$CDM)~\cite{riess98,perlmutter99}. Such a model provides the best description of cosmological observations such as temperature fluctuations of the Cosmic Microwave Background (CMB)~\cite{planck18}, luminosity distances to Type Ia Supernovae (SNe)~\cite{scolnic18}, large-scale clustering of galaxies (LSS), and weak gravitational lensing (WL)~\cite{sdss21, des21a, des21b, kids21}. Despite its tremendous success, this model presents theoretical caveats, such as the value of the vacuum energy density~\cite{Weinberg:1988cp,Padmanabhan:2002ji}, in addition to observational challenges e.g. the $\simeq 5\sigma$ Hubble constant tension between CMB and SNe observations~\cite{divalentino21, shah21, riess21}, as well as milder tensions between matter density perturbation estimates from CMB and LSS, and slightly enhanced CMB lensing amplitude than predicted by the $\Lambda$CDM model. These conflicting measurements may hint at physics beyond the standard cosmology.

Given the necessity to probe the Universe at larger and deeper scales, cosmological surveys like Javalambre-Physics of the Accelerated Universe Astrophysical Survey (J-PAS)~\cite{jpas14,minijpas21a}, Dark Energy Spectroscopic Instrument (DESI)~\cite{desi16}, Euclid~\cite{euclid18}, Square Kilometer Array (SKA)~\cite{ska20} and the Large Synoptic Survey Telescope (LSST)~\cite{lsst18} were proposed and developed. They will improve the constraints we currently have on the parameters of the $\Lambda$CDM model, and probe departures from it with unprecedented sensitivity. In order to extract the most of cosmological information from the tantalising amount of data to come, the deployment of machine learning (ML) algorithms on Physics and Astronomy~\cite{ML19a, ML19b} is becoming crucial to accelerate data processing and improve statistical inference. Some recent applications of ML on Cosmology focuses on reconstructing the late-time cosmic expansion history to test fundamental hypothesis of the standard model and constrain its parameters~\cite{li19, liu19, wu19, arjona20, escamilla-rivera20, wang20a, wang20b, wang21, liu21, garcia22, dialektopoulos22, mukherjee22, gomez-vargas23, tonghua23}, cosmological model discrimination with LSS and WL~\cite{agarwal12, agarwal14, ravanbakhsh17, merten19, peel19, ribli19, fluri19, pan19, ntampaka19, matilla20, villaescusa-navarro22}, predicting structure formation~\cite{kamdar16a, kamdar16b, lucie-smith18, he18, lucie-smith19, ramanah19, tsizh20, murakami20, chacon21, vonMarttens22, piras22}, probing the era of reionisation~\cite{hassan18, gillet19, chardin19, LaPlante19, hassan20, mangena20, prelogovic21}, photometric redshift estimation~\cite{collister03, hogan15, sadeh16, bilicki18, gomes18, desprez21, cabayol21, kunsagi-mate22}, besides the classification of astrophysical sources~\cite{kurcz16, kim17, beck19, minijpas21b} and transient objects~\cite{lochner16, muthukrishna19a, muthukrishna19b, fremling21}. These analyses reveal that ML algorithms are able to recover the underlying cosmology from data and simulations with greater precision than traditionally used techniques e.g. 2-point correlation function and power spectrum, in addition to Markov Chain Monte Carlo (MCMC) methods.

In this paper we discuss the ability to measure the Hubble Constant $H_0$ from cosmic chronometers  measurements ($H(z)$) using different ML algorithms. We first produce $H(z)$ synthetic data-sets with different number of data points and measurement uncertainties, in order to perform a benchmark test of the $H_0$ constraints for each algorithms given the quality of the input data. Rather than performing a numerical reconstruction across the redshift range probed by the data, and then fitting $H_0$, we carry out an {extrapolation} of the reconstructed $H(z)$ values down to $z=0$.  We also compare their performance with other non-parametric reconstruction methods, such as the {popularly adopted} Gaussian Processes (GAP)~\cite{seikel12}. Our goal is to verify whether they can provide a competitive cross-check with the GAP. 

{The paper is structured as follows: Section 2 is dedicated to the cosmological framework and the simulations produced for our analysis. Section 3 explains how this analysis is performed, along with the metrics adopted for algorithm performance evaluation. Section 4 presents our results; finally our main conclusions and final remarks are presented in Section 5.} 


\section{Simulations}

\begin{table}
	\centering \caption{31 CC $H(z)$ measurements obtained from the differential age method used in our analysis}\label{tab:Hz}
	\begin{tabular}{ccc}
		\hline
		\hline
		z 		& $H(z)$ (km $\rm s^{-1}$ $\rm Mpc^{-1}$) & References \\
		\hline
		0.09 	&	$69\pm12$		&   \cite{jimenez03} \\
		\hline
		0.17	&	$83\pm8$		&	\\
		0.27	&	$77\pm14$		&	\\
		0.4		&	$95\pm17$		&	\\
		0.9		&	$117\pm23$		&   \cite{simon05} \\
		1.3		&	$168\pm17$		&	\\
		1.43	&	$177\pm18$		&	\\
		1.53	&	$140\pm14$		&	\\
		1.75	&	$202\pm40$		&	\\
		\hline
		0.48	&	$97\pm62$		&   \cite{stern10} \\
		0.88	&	$90\pm40$		&	\\
		\hline
		0.1791	&	$75\pm4$		&	\\
		0.1993	&	$75\pm5$		&	\\
		0.3519	&	$83\pm14$		&	\\
		0.5929	&	$104\pm13$		&	\cite{moresco12} \\
		0.6797	&	$92\pm8$		&	\\
		0.7812	&	$105\pm12$		&	\\
		0.8754	&	$125\pm17$		&	\\
		1.037	&	$154\pm20$		&	\\
		\hline
		0.07	&	$69\pm19.6$		&	\\
		0.12	&	$68.6\pm26.2$	&	\cite{zhang14} \\
		0.2		&	$72.9\pm29.6$	&	\\
		0.28	&	$88.8\pm36.6$	&	\\
		\hline
		1.363	&	$160\pm33.6$	&	\cite{moresco15} \\
		1.965	&	$186.5\pm50.4$	&	\\
		\hline
		0.3802	&	$83\pm13.5$		&	\\
		0.4004	&	$77\pm10.2$		&	\\
		0.4247	&	$87.1\pm11.2$	&	\cite{moresco16} \\
		0.44497	&	$92.8\pm12.9$	&	\\
		0.4783	&	$80.9\pm9$		&	\\
		\hline
		0.47    &   $89\pm49.6$     &  \cite{ratsimbazafy17} \\
		\hline
	\end{tabular}
\end{table}

\subsection{Prescription}

In order to compare how different predicting algorithm perform with different quality of data, we produce sets of $H(z)$ simulated data sets and adopt the following prescription:

\textbf{(i)} We assume as fiducial cosmology the flat $\Lambda$CDM model given by Planck 2018 (TT, TE, EE+lowE+lensing; hereafter P18)~\cite{planck18}: 
\begin{eqnarray}\label{eq:model}
H^{\rm fid}_0 &=& 67.36 \pm 0.54 \, \mathrm{km \, s}^{-1} \, \mathrm{Mpc}^{-1} \,\\ 
\Omega^{\rm fid}_{\rm m} &=& 0.3166 \pm 0.0084 \,\\
\Omega^{\rm fid}_{\Lambda} &=& 1-\Omega_{\rm m} \,,
\end{eqnarray}
so that the Hubble parameter follows the Friedmann equation for the fiducial $\Lambda$CDM model
\begin{equation}\label{eq:hz_fid}
\left[\frac{H^{\rm fid}(z)}{H^{\rm fid}_0}\right]^2 = \Omega^{\rm fid}_{\rm m}(1+z)^3 + \Omega^{\rm fid}_{\Lambda} \,.
\end{equation}

\textbf{(ii)} We compute the values of $H(z)$ considering the $N_z$ data points following a redshift distribution $p(z)$ such as~\cite{wang20a}
\begin{equation}\label{eq:pz}
p(z; \; k,\theta) = z^{k-1}\frac{e^{-z/\theta}}{\theta^{k}\Gamma(k)} \;,
\end{equation}
where we fix $\theta$ and $k$ to their respective best fits to the real cosmic chronometers data, as in~\cite{wang20a}, i.e., $\theta_{\rm bf}=0.647$ and $k=1.048$. 

\textbf{(iii)} In order to understand how our knowledge of $H(z)$ along the redshift space affects the performance of the statistical learning, we provide different sets of $H(z)$ assuming different numbers of points $N_z$ = $20$, $30$, $50$ and $80$, and assuming different relative uncertainties values, i.e., $\sigma_H/H=0.008, 0.01, 0.03, 0.05, 0.08$. This variation of $N_z$ and $\sigma_H(z)$ allows to evaluate what level of accuracy of measurements of $H(z)$ is necessary in order to obtain a specific precision on the prediction of $H_0$.

\textbf{(iv)} We also produce $H(z)$ simulations based on the current cosmic chronometer data, which consists of $N_z=31$ measurements presented in~\ref{tab:Hz} - see also Table I in~\cite{wang20a}.

Such a prescription provides a benchmark to test the performance of the ML algorithms deployed.  

\subsection{Uncertainty estimation}

Although these algorithms are able to provide measurements of $H_0$ at a given redshift, they do not provide their uncertainties. We develop a Monte Carlo-bootstrap (MC-bootstrap) method for this purpose, described as follows

\begin{itemize}

\item Rather than creating a single simulation centered on the fiducial model for each data-set (item {\bf (i)} of section II), we produce $H(z)$ measurements at a given redshift following a normal distribution centred around its fiducial value according to $\mathcal{N}(H^{\rm fid}(z),\sigma_{H}/H)$. $H^{\rm fid}(z)$ represents the $H(z)$ value given by the fiducial Cosmology, whereas $\sigma_{H}$ consists on its uncertainty as described in the item {\bf (iii)} of section II. 

\item As for the "real data" simulations, described in item {\bf (iv)} of section II, we replace the $i$-th $H(z)$ measurement presented in the second column of Table~\ref{tab:Hz} by a value drawn from a normal distribution centered on the fiducial model, i.e, $\mathcal{N}(H^{\rm fid}(z_i),\sigma_{H;i})$, where $z_i$ represents the redshift of each data point and $\sigma_{H;i}$ its corresponding uncertainty - first and third columns in~\ref{tab:Hz}, respectively.

\item We repeat this procedure 100 times for each data-set of $N_z$ data-points with $\sigma_H/H$ uncertainties as described in the item {\bf (iii)} and {\bf (iv)} of section II. 

\item The 100 MC realisations produced for each case are provided as inputs for each ML algorithms described in subsection IIIa 

\item We report the average and standard deviation of these 100 values as the $H_0$ measurement and uncertainty, respectively, for each $N_z$ and $\sigma_H/H$ case. Same applies for the "real data" simulations.

\end{itemize}

\section{Analysis}

\subsection{Methods}

Our regression analysis are carried out on all simulated and "real" data-sets with several ML algorithms available in the \texttt{scikit-learn} package~\footnote{\url{https://scikit-learn.org/stable/}}\cite{sklearn11}. Firstly we divide our input sample into training and testing data-sets as
\begin{lstlisting}
z_train,z_test,hz_train,hz_test =
train_test_split(z,hz,test_size=0.25,
random_state=42)
\end{lstlisting}
so that our testing sub-set contains 25\% of the original sample size. Then we deploy different ML algorithms on the training test, looking for the ``best combination" of hyperparameters with the help of \textit{GridSearchCV}\footnote{\url{https://scikit-learn.org/stable/modules/generated/sklearn.model_selection.GridSearchCV.html}}. This function of \texttt{scikit-learn}, given a ML method, performs the learning with 
all the combination of hyperparameters and shows the performance of every combination - or each one of them - during the cross-validation (CV) procedure~\footnote{\url{https://scikit-learn.org/stable/modules/cross_validation.html}}. Such a procedure is performed for the sake of avoiding overfitting on the test set. We chose CV $=3$ in our analysis, given the limited number of $H(z)$ data-points.

The ML methods deployed in our analysis are given as follows:

\begin{itemize}

\item {\bf Extra-Trees (EXT)}: An ensemble of randomised decision trees (extra-trees). The goal of the algorithm is to create a model that predicts the value of a target variable by learning simple decision rules inferred from the data features. A tree can be seen as a piecewise constant approximation~\footnote{\url{https://scikit-learn.org/stable/modules/tree.html}}\cite{BRE}. We evaluate the algorithm hyperpameter values that best fit the input simulations through a grid search. Hence, our grid search over the EXT hyperpameters are given by:

\begin{lstlisting}
gcv = GridSearchCV(ExtraTreesRegressor
(min_samples_split=2,
random_state=42),
param_grid={
'n_estimators': np.arange(1,100,2),
'max_depth': np.arange(1,10,2),
cv=3, refit=True)
\end{lstlisting}

\item {\bf Artificial Neural Network (ANN)}: A Multi-layered Perceptron algorithm that trains using backpropagation with no activation function in the output layer\footnote{\url{https://scikit-learn.org/stable/modules/neural_networks_supervised.html}}\cite{Rumelhart}. The ANN hyperparameter grid search consists of:

\begin{lstlisting}
gcv = GridSearchCV(MLPRegressor
({activation='relu'},
solver='lbfgs', 
learning_rate='adaptive', 
max_iter=200, 
random_state=42),
param_grid={
'hidden_layer_sizes': np.arange(10,250,10),
},
cv=3, refit=True)
\end{lstlisting}

\item {\bf Gradient Boosting Regression (GBR)}: This estimator builds an additive model in a forward stage-wise fashion; it allows for the optimisation of arbitrary differentiable loss functions. In each stage a regression tree is fit on the negative gradient of the given loss function
\footnote{\url{https://scikit-learn.org/stable/modules/generated/sklearn.ensemble.GradientBoostingRegressor.html\#sklearn.ensemble.GradientBoostingRegressor}}\cite{Friedman01}. The grid search over the GBR hyperparameters corresponds to:

\begin{lstlisting}
gcv = GridSearchCV(
GradientBoostingRegressor(random_state=42),
param_grid={
'n_estimators': np.arange(1,200,5),
},
cv=3, refit=True)
\end{lstlisting}

\item {\bf Support Vector Machines (SVM)}: A linear model that creates a line or hyperplane to separate data into different classes \cite{bishop,smola}. Originally developed for classification problems, it was also extended for regression, as in the goal of this work\footnote{\url{https://scikit-learn.org/stable/modules/svm.html}}. The hyperparameter grid search of the SVM method reads

\begin{lstlisting}
gcv = GridSearchCV(SVR(kernel='poly', 
C=100,
gamma='auto', epsilon=.1,
coef0=1),
param_grid={
'degree': np.arange(1, 10),},
cv=3, refit=True)
\end{lstlisting}

\end{itemize}

Note that we adopted the default evaluation metric for each ML algorithm as defined by the \texttt{scikit-learn} package. So the EXT method uses the squared error metric to define the quality of the tree split - likewise for the GBR and ANN loss functions - whereas the SVM method assumes $\epsilon=0.1$, so that samples whose prediction is at least $\epsilon$ away from their true target are penalised\footnote{\url{https://scikit-learn.org/stable/modules/svm.html\#mathematical-formulation}}.

In order to evaluate the performance of these methods, we report the results of the training and test score as
\begin{lstlisting}
print(gcv.best_estimator_.predict([[0.]]),
gcv.score(z_train,hz_train), 
gcv.score(z_test,hz_test))
\end{lstlisting}

Moreover, we deploy the well-known Gaussian Processes regression (GAP) algorithm on the same simulated data-sets using the {\sc GaPP} package~\cite{seikel12}. We compare the results obtained with the ML algorithms just described with GAP since the latter has been widely used in the literature for similar purposes for about a decade. Two GAP kernels are assumed in our analysis, namely the Squared Exponential (SqExp) and Mat\'ern(5/2) (Mat52). We justify these choices on the basis that the SqExp kernel exhibits greater differentiability than the Mat52, which may result in a larger degree of smoothing on the data reconstruction - hence, smaller reconstruction uncertainties - that may or may not fully represent the underlying data.

\subsection{Robustness of results}

We define the bias (b) as the average displacement of the predicted Hubble Constant ($H^{\rm pred}_0$), obtained from the MC-bootstrap method, from the fiducial value, i.e., $\Delta H_0=H^{\rm pred}_0 - H^{\rm fid}_0$, and the Mean Squared Error (MSE) as the average squared displacement:
\begin{equation}
    \mathrm{b} = \langle \Delta H_0 \rangle, ~~~ \mathrm{MSE} = \langle \Delta H_0^2 \rangle \,.
\end{equation}
Using the definition of variance,  we estimate the bias-variance tradeoff of our analysis
\begin{equation}
    \mathrm{BVT} = \langle \Delta H_0^2 \rangle - \langle \Delta H_0 \rangle^2 = \mathrm{MSE} - \mathrm{b}^2 \;,
\end{equation}
therefore we can evaluate the performance of these algorithms for each simulated data-set specification.

\section{Results}\label{sec:results} 

\begin{figure*}[]
\includegraphics[width=0.49\textwidth, height=7.3cm]{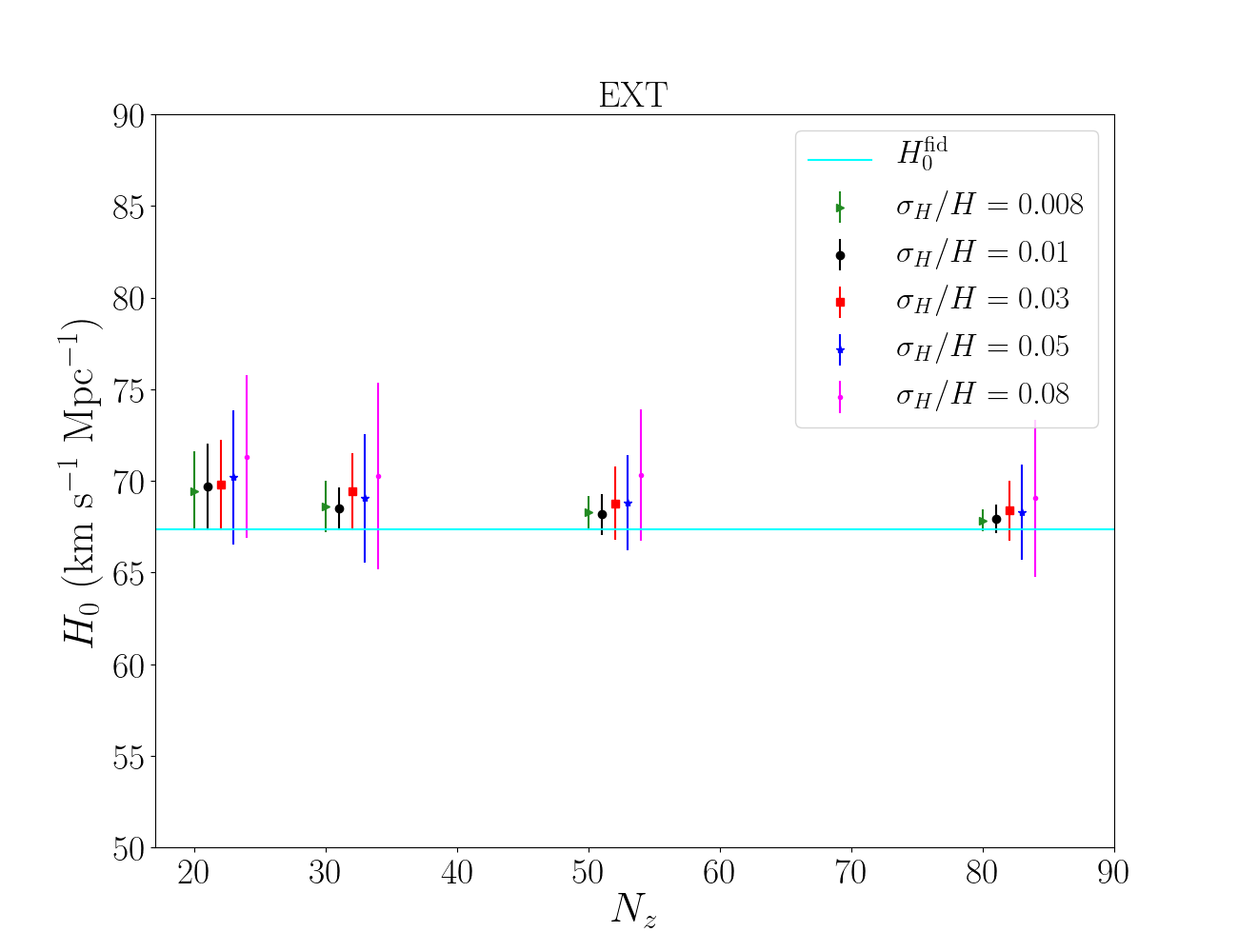}
\includegraphics[width=0.49\textwidth, height=7.3cm]{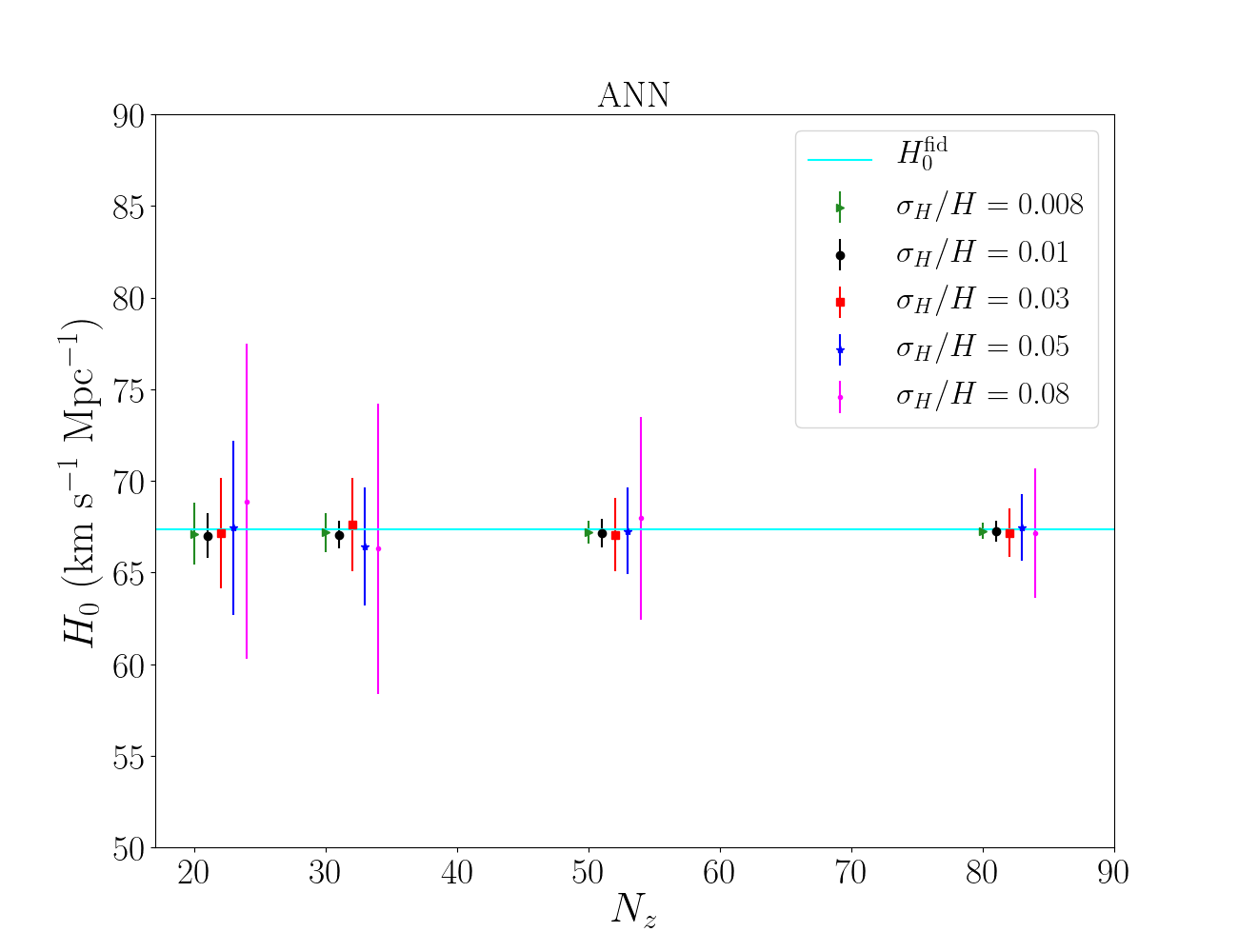}
\includegraphics[width=0.49\textwidth, height=7.3cm]{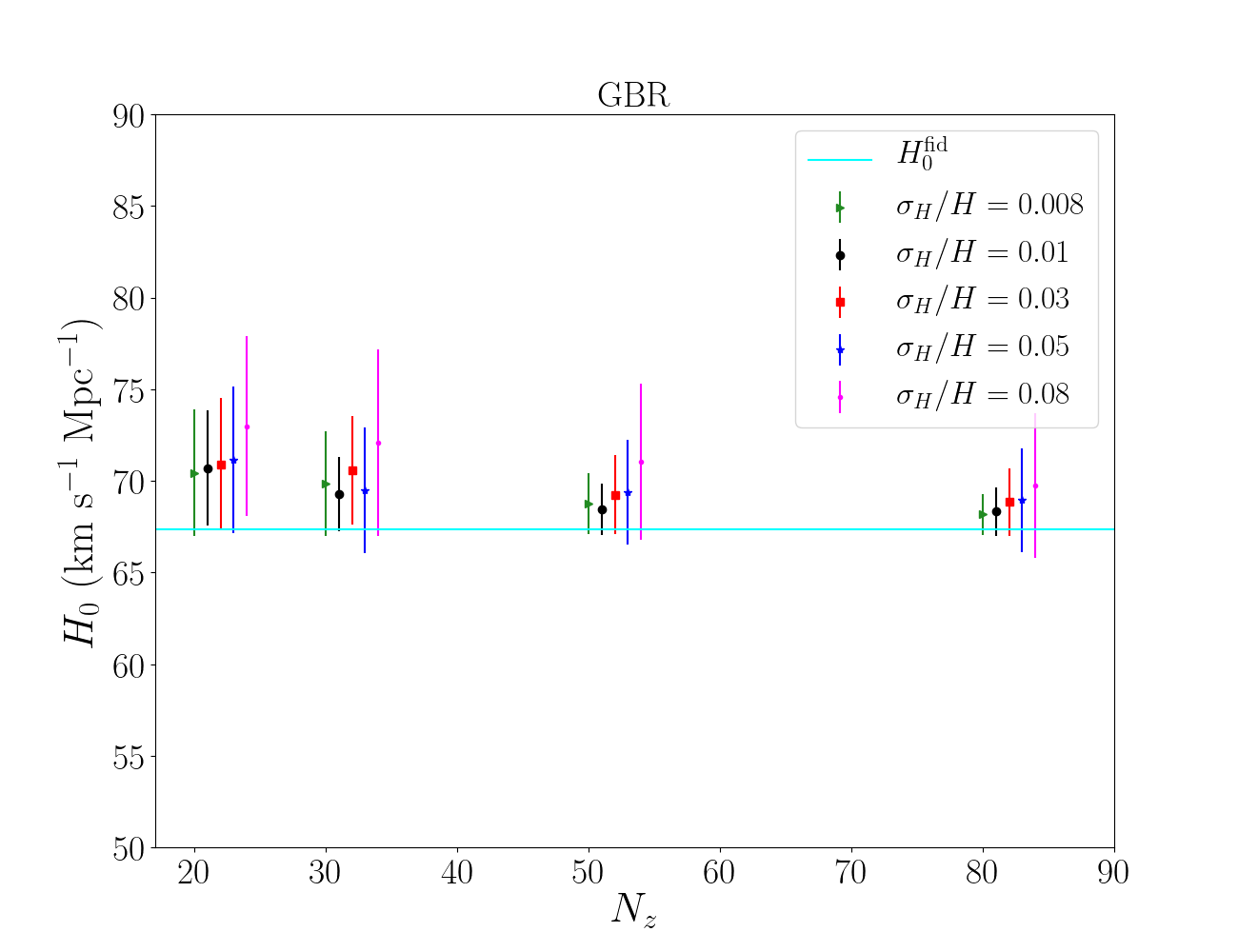}
\includegraphics[width=0.49\textwidth, height=7.3cm]{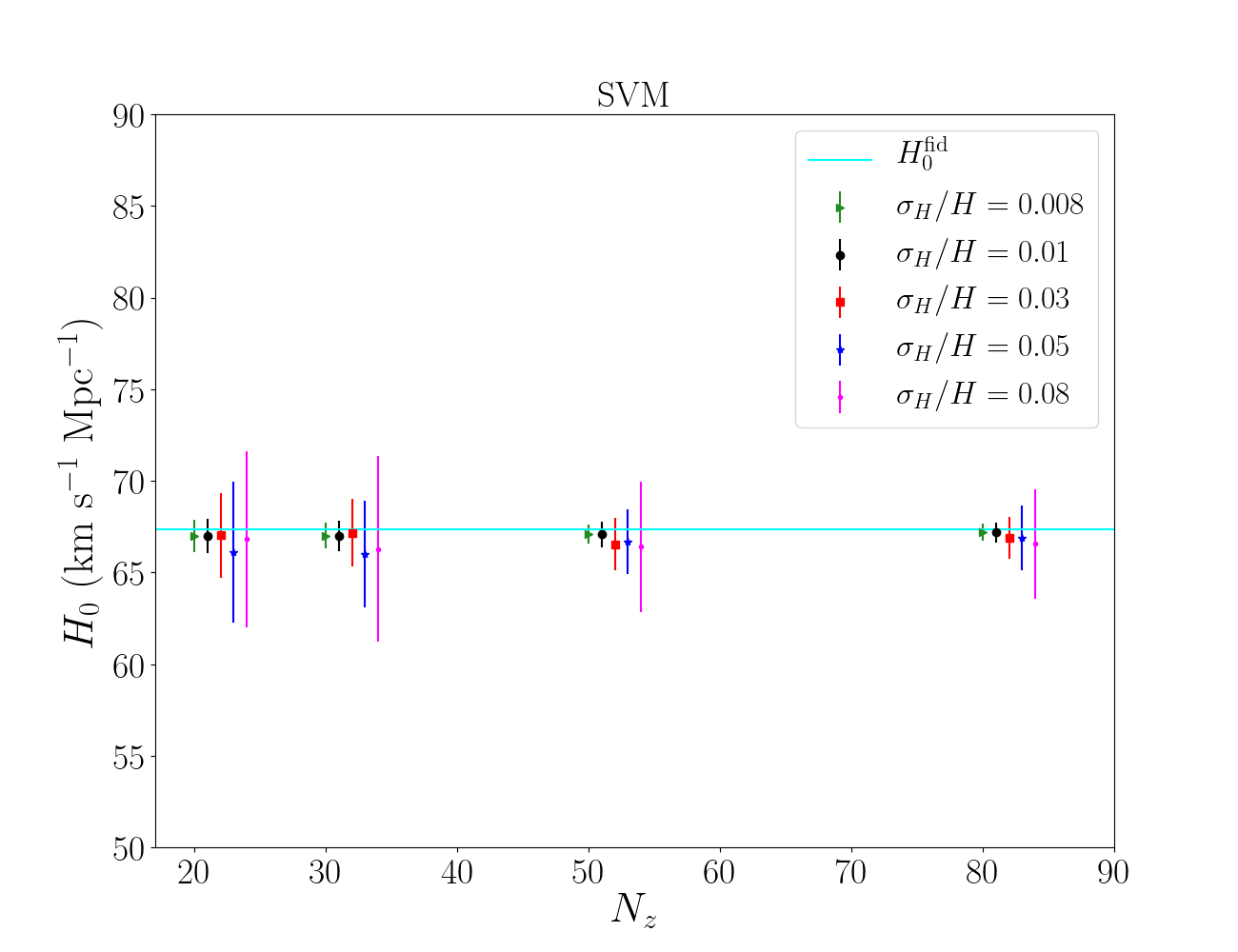}
\includegraphics[width=0.49\textwidth, height=7.3cm]{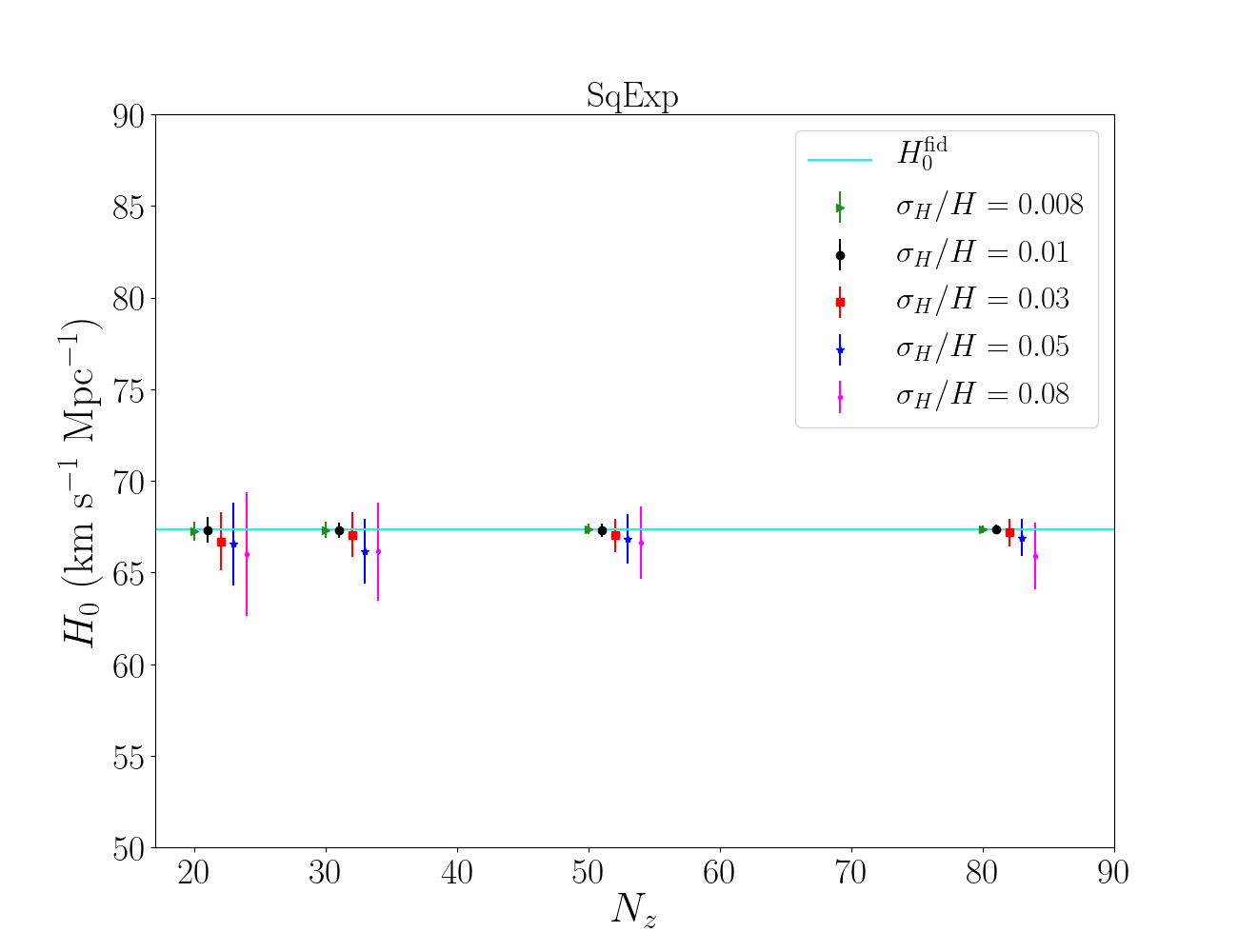}
\includegraphics[width=0.49\textwidth, height=7.3cm]{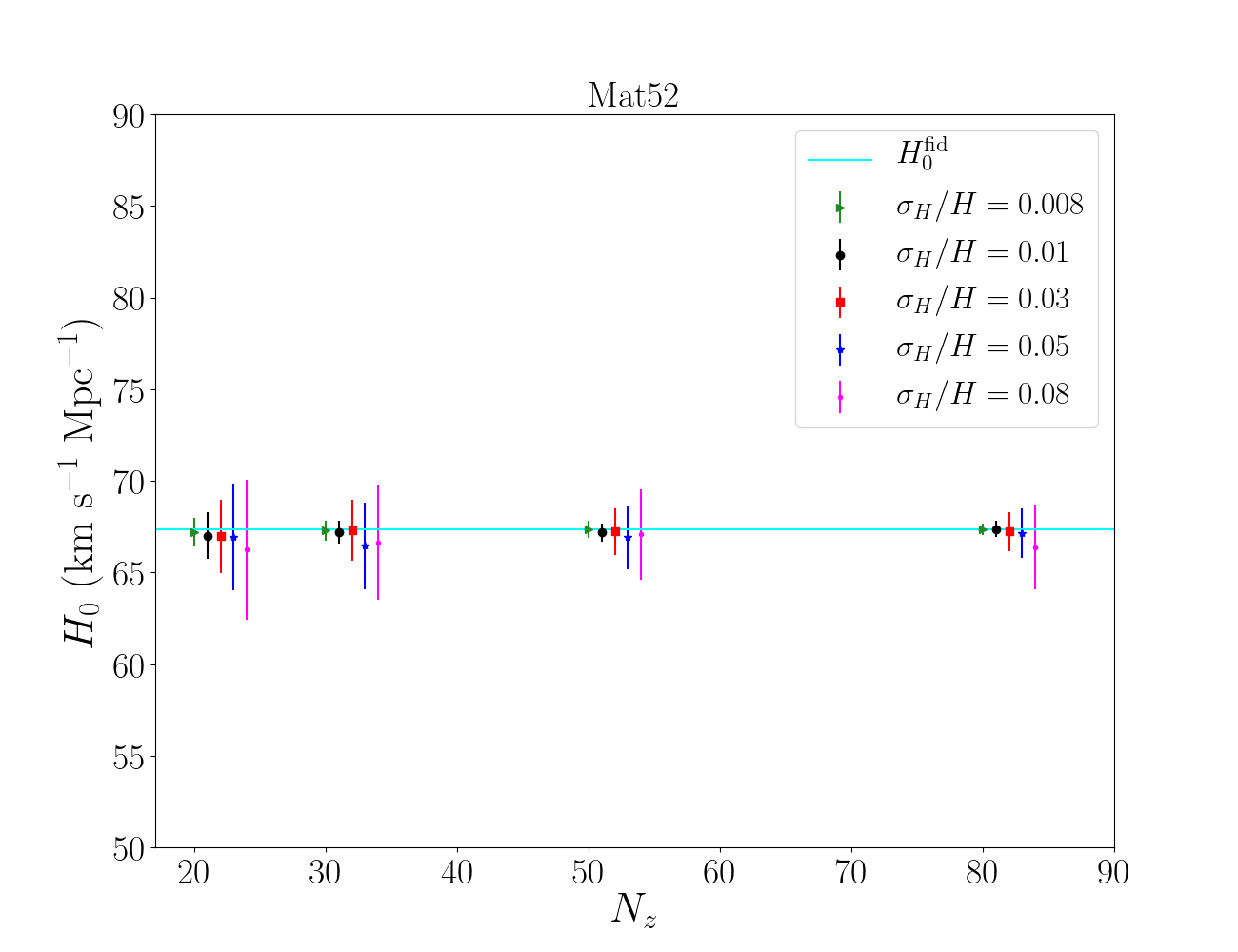}
\caption{$H_0$ measurements from the algorithm EXT (top left), ANN (top right), GBR (center left), SVM (center right), SqExp (lower left) and Mat52 (lower right), plotted against the number of simulated $H(z)$ measurements. Each data point represents different $\sigma_H/H$ values, whereas the light blue horizontal lines denote the fiducial $H_0$ value.}
\label{fig:H0_measurements}
\end{figure*}

\begin{figure*}[]
\includegraphics[width=0.49\textwidth, height=7.3cm]{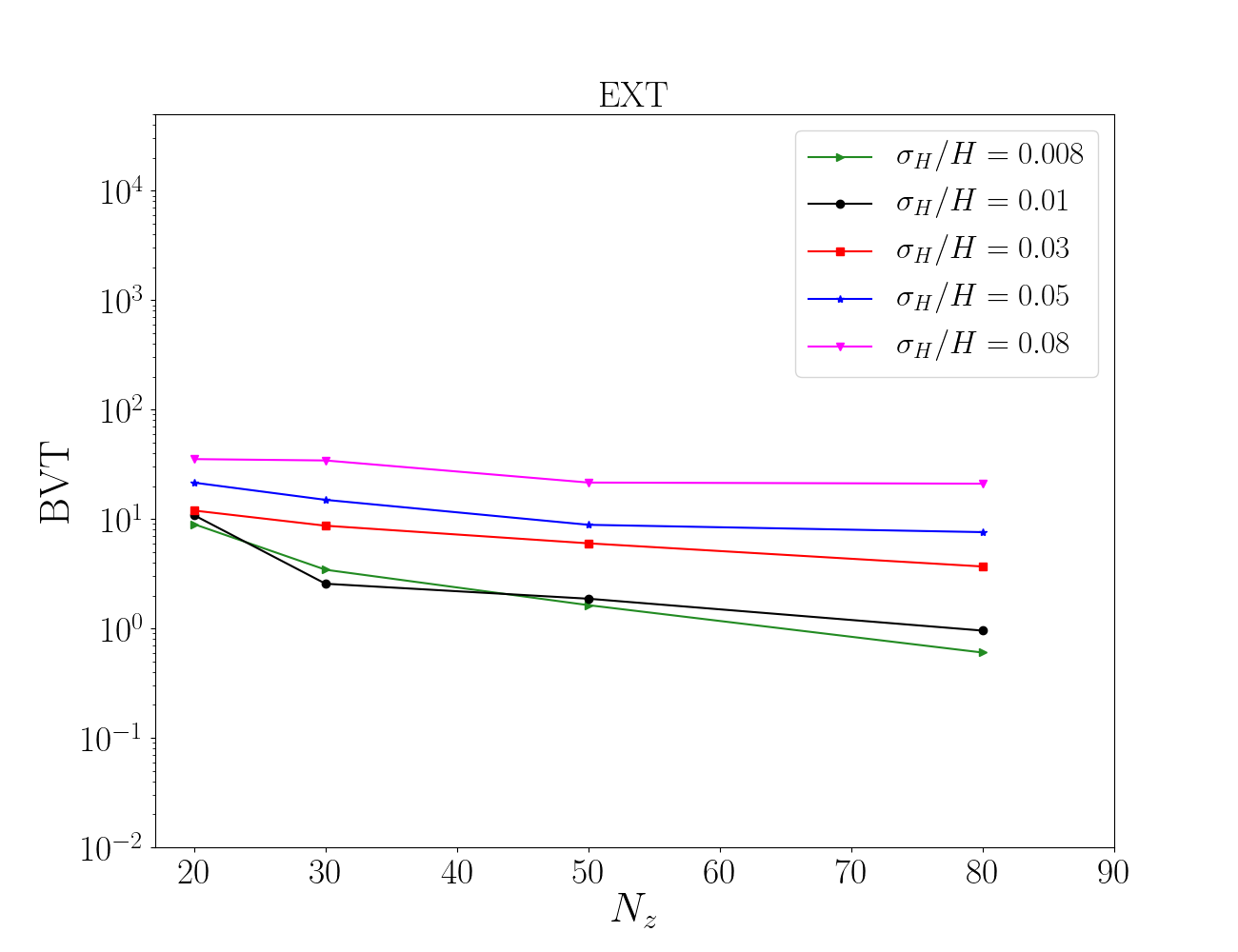}
\includegraphics[width=0.49\textwidth, height=7.3cm]{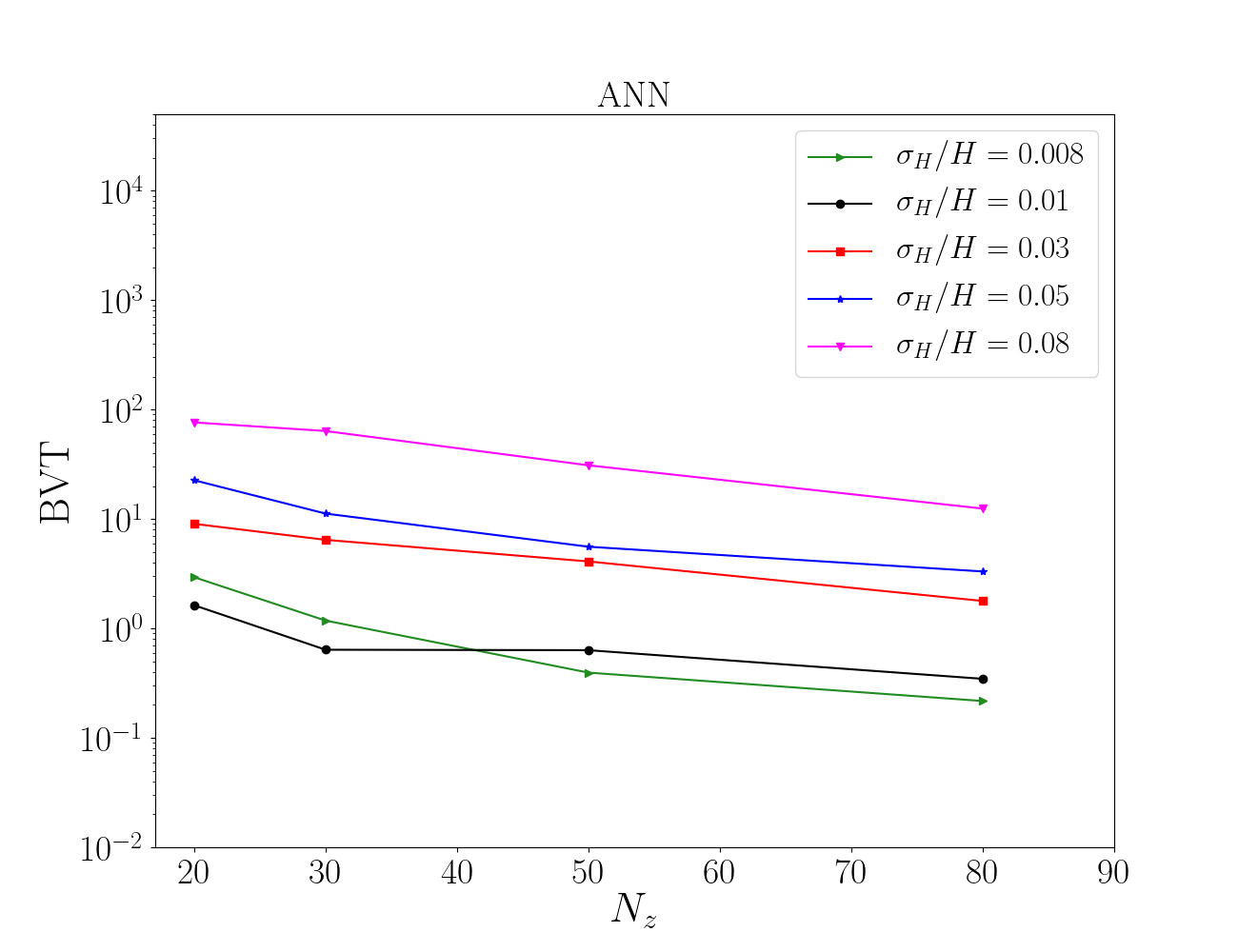}
\includegraphics[width=0.49\textwidth, height=7.3cm]{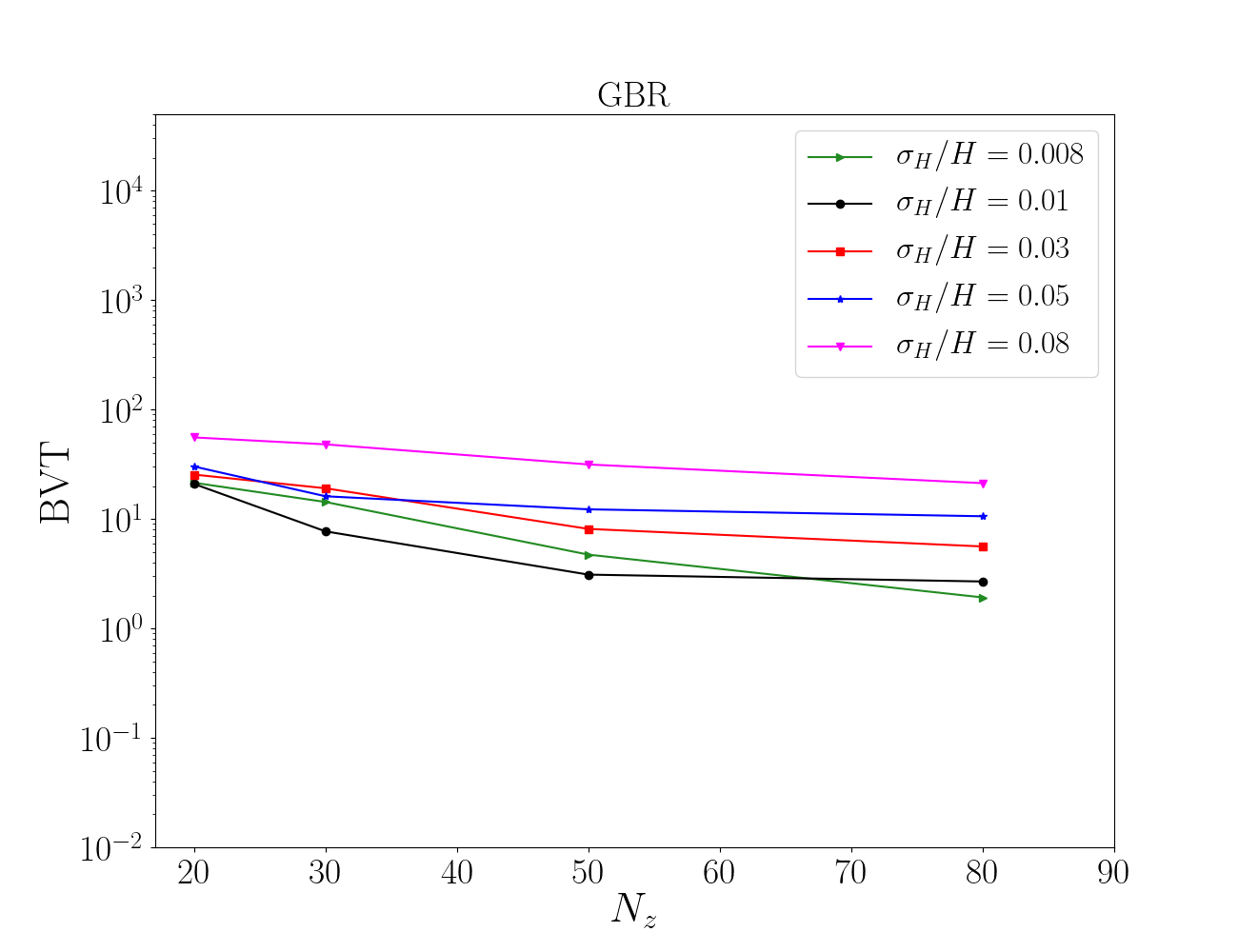}
\includegraphics[width=0.49\textwidth, height=7.3cm]{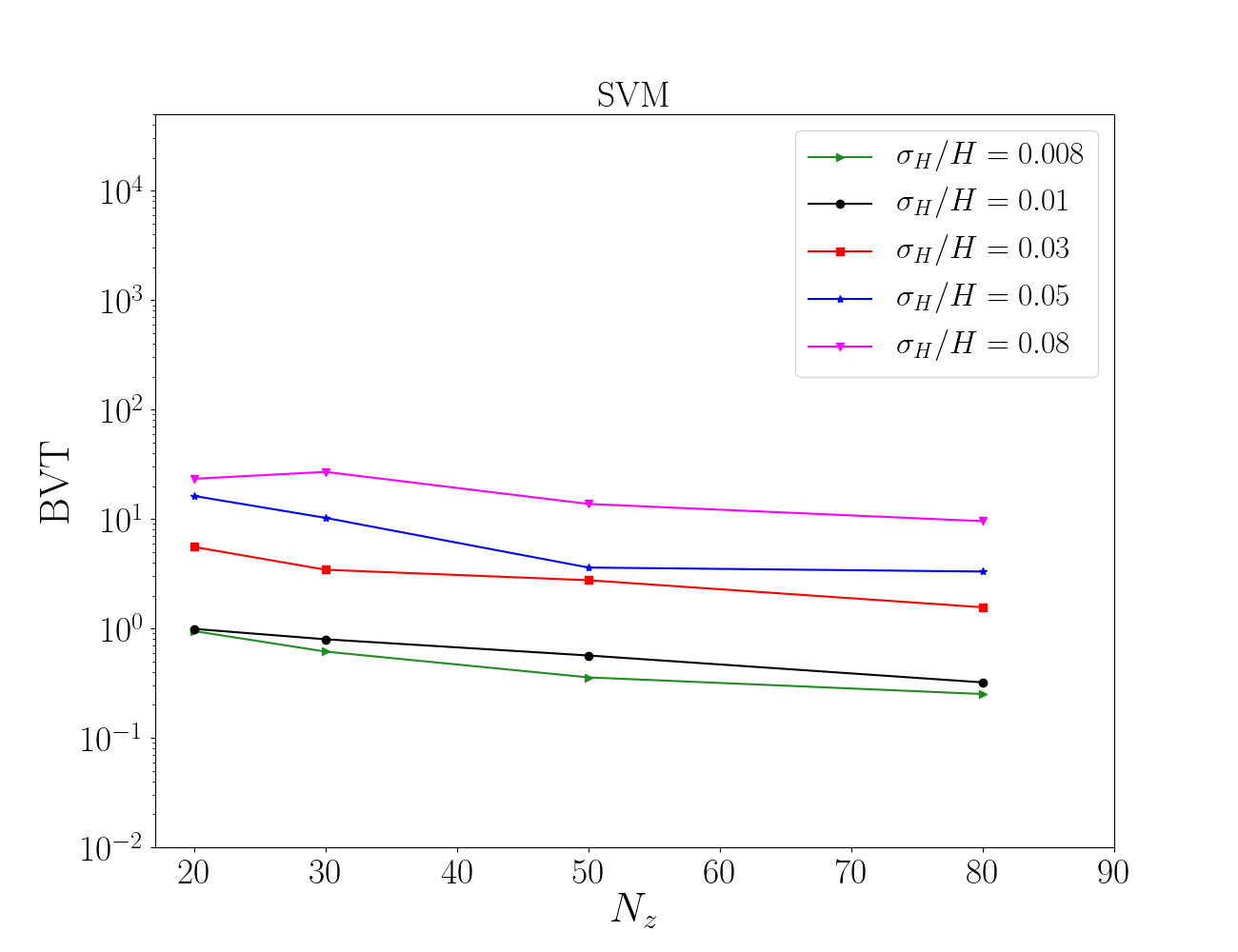}
\includegraphics[width=0.49\textwidth, height=7.3cm]{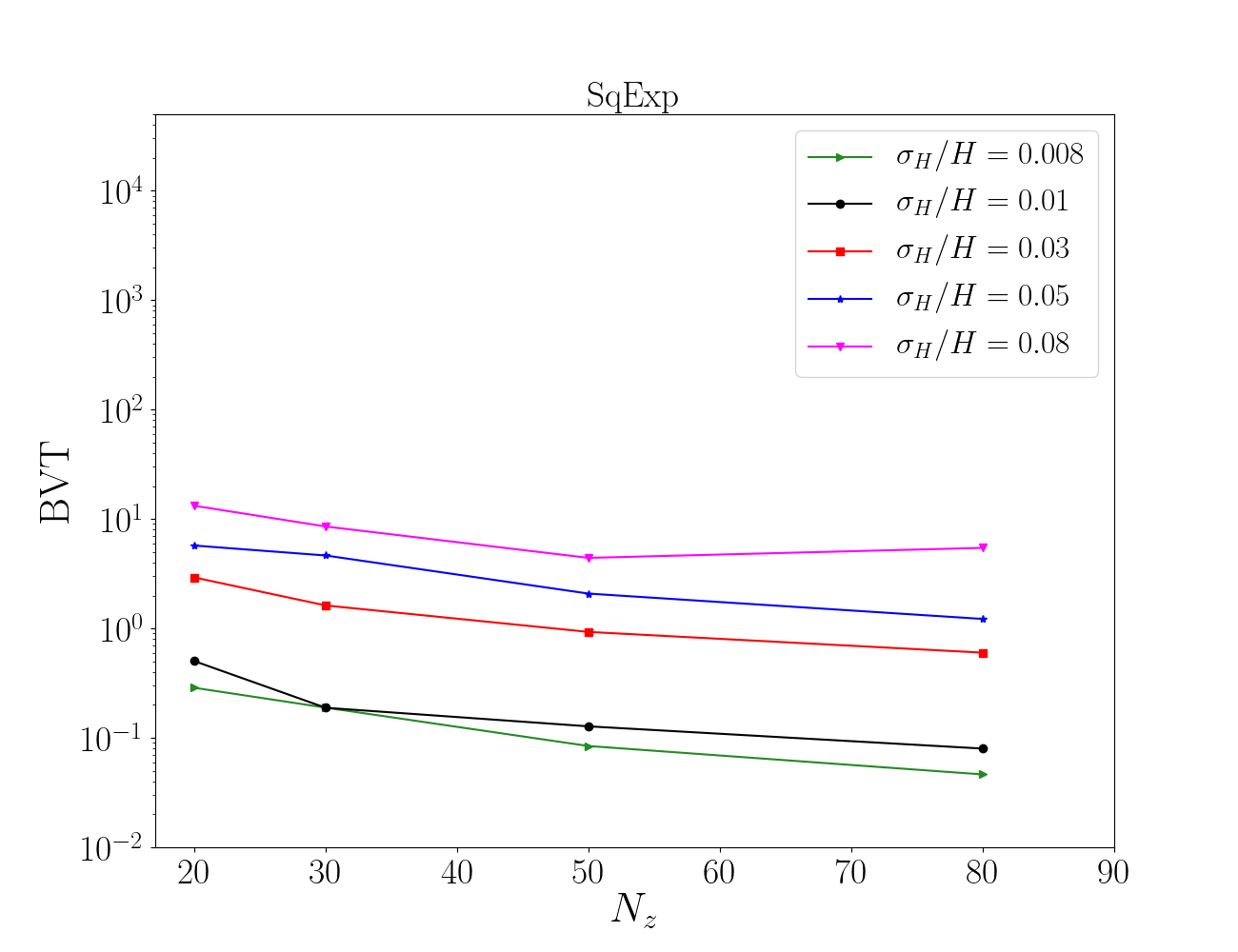}
\includegraphics[width=0.49\textwidth, height=7.3cm]{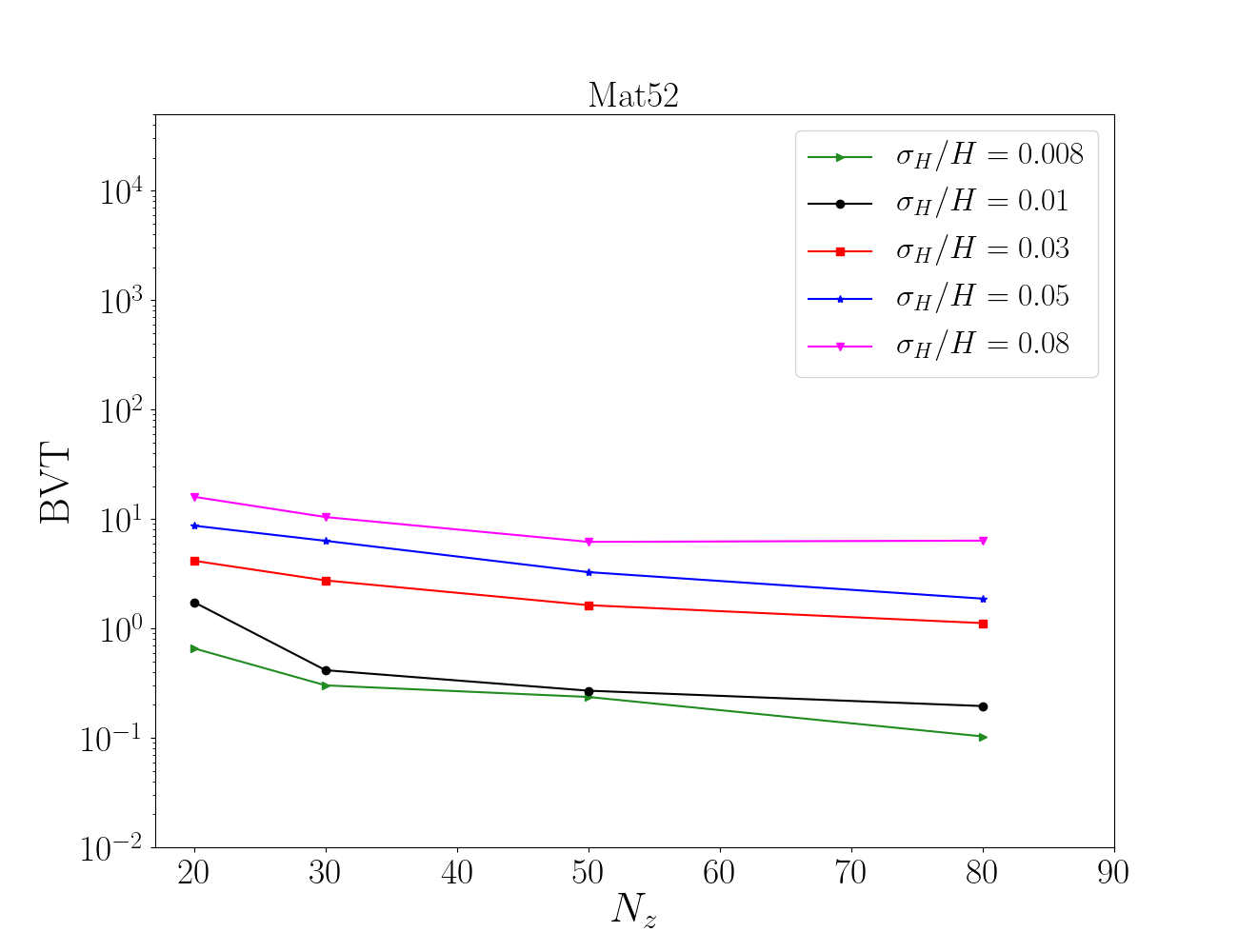}
\caption{Same as Fig.~\ref{fig:H0_measurements}, but for the BVT. Each data point corresponds to different $\sigma_H/H$ values.}
\label{fig:BVT_measurements}
\end{figure*}

\begin{figure*}[]
\includegraphics[width=0.49\textwidth, height=7.3cm]{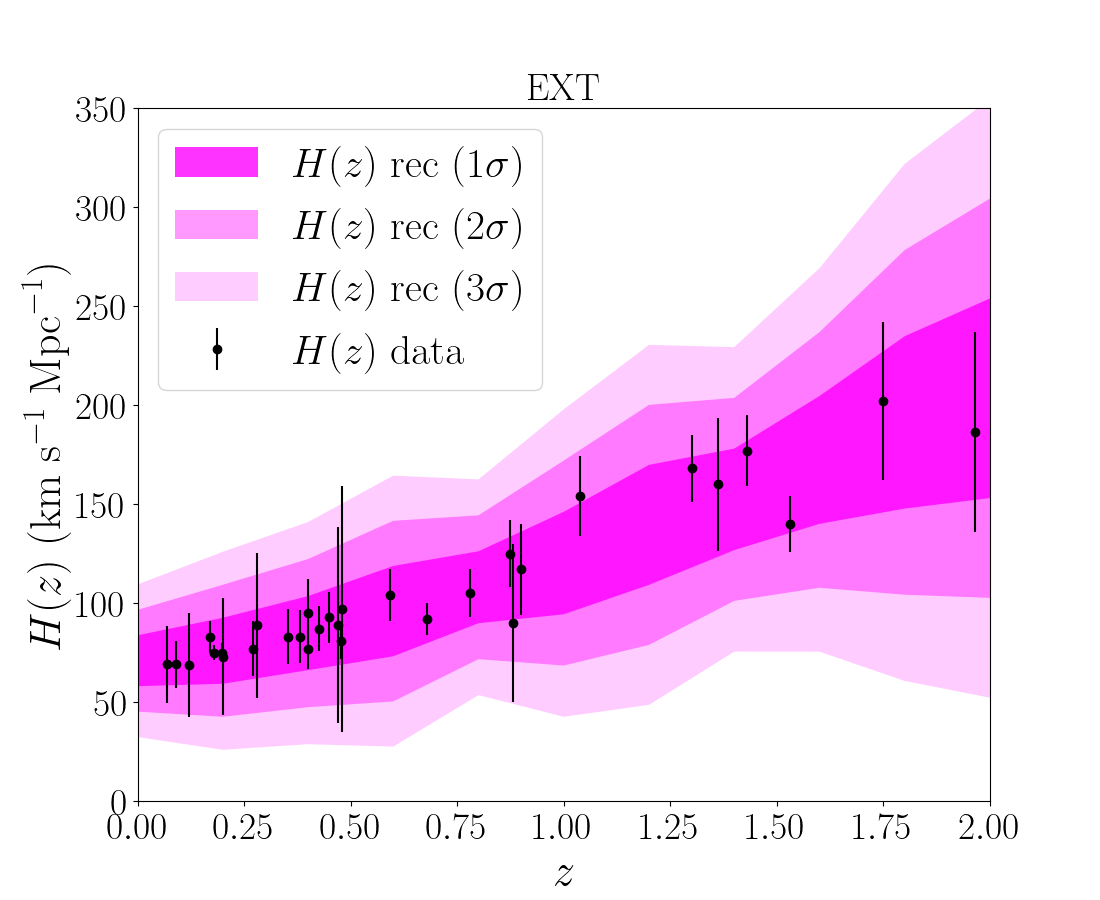}
\includegraphics[width=0.49\textwidth, height=7.3cm]{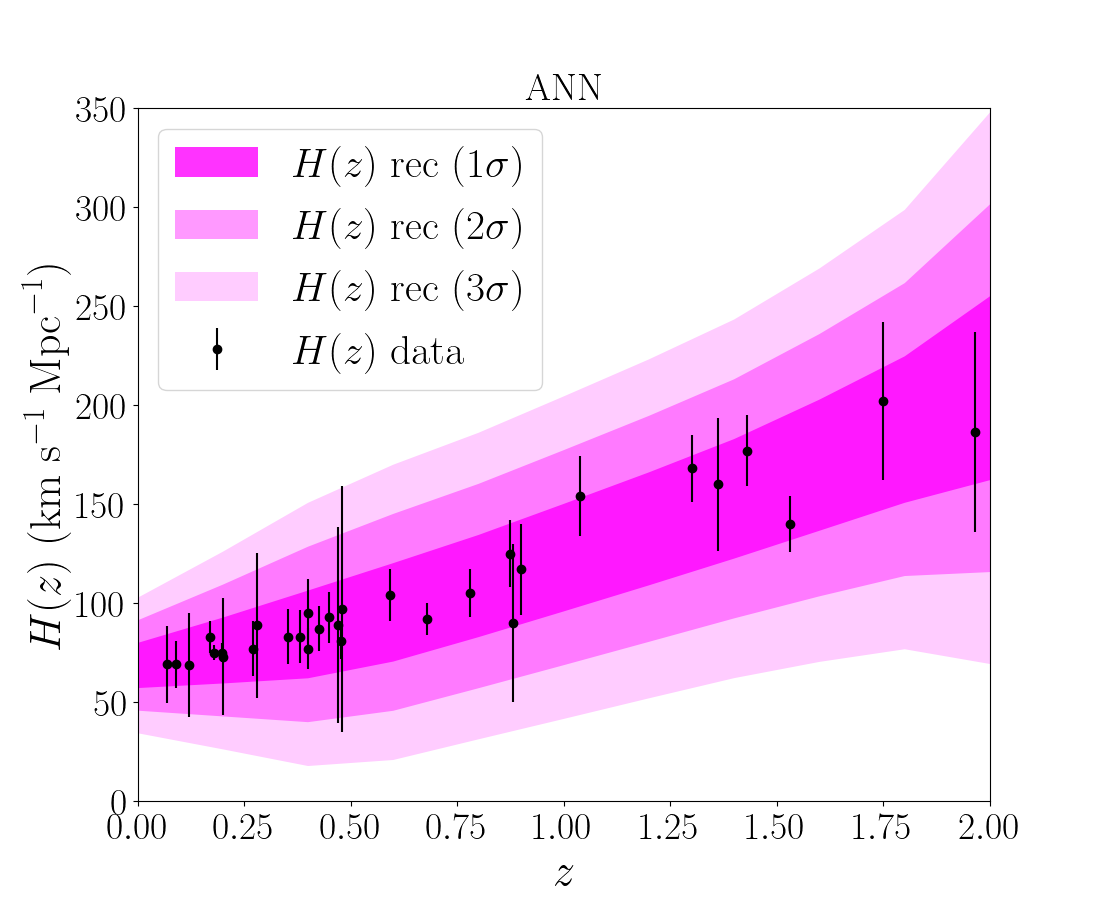}
\includegraphics[width=0.49\textwidth, height=7.3cm]{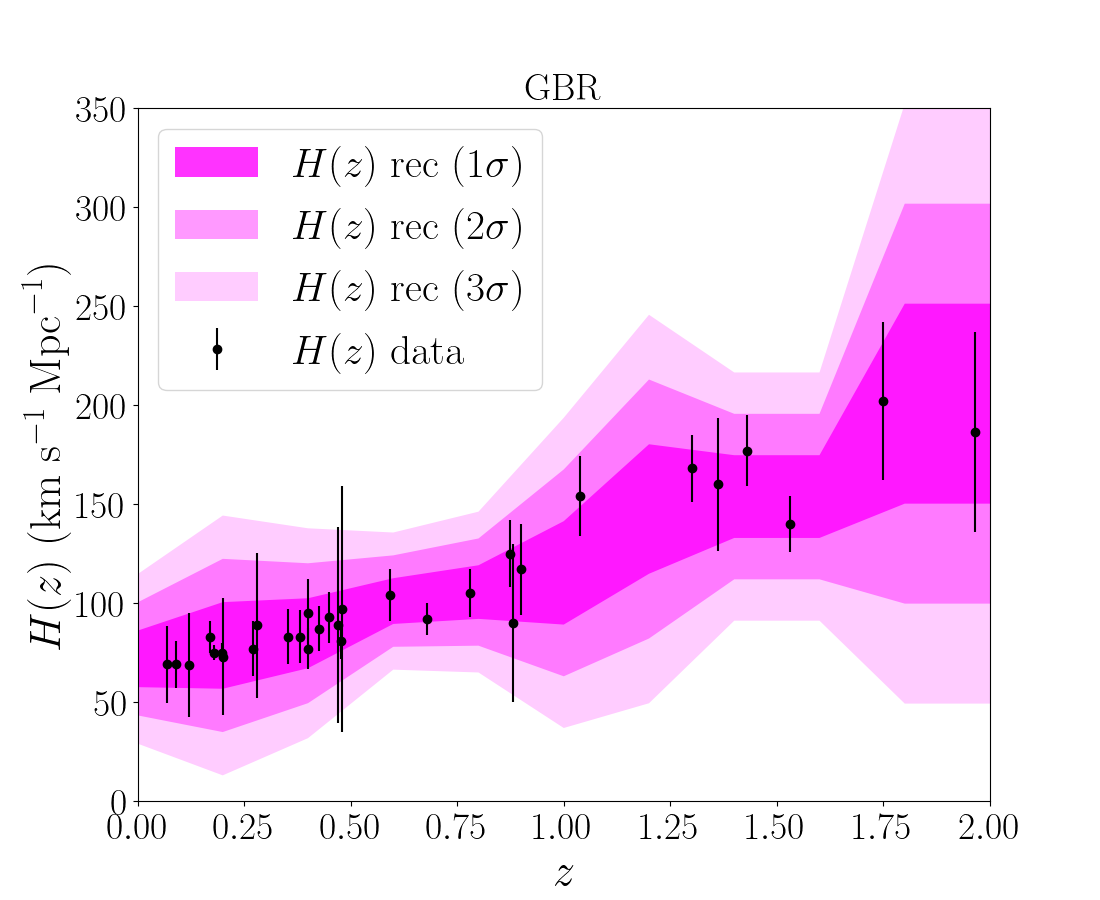}
\includegraphics[width=0.49\textwidth, height=7.3cm]{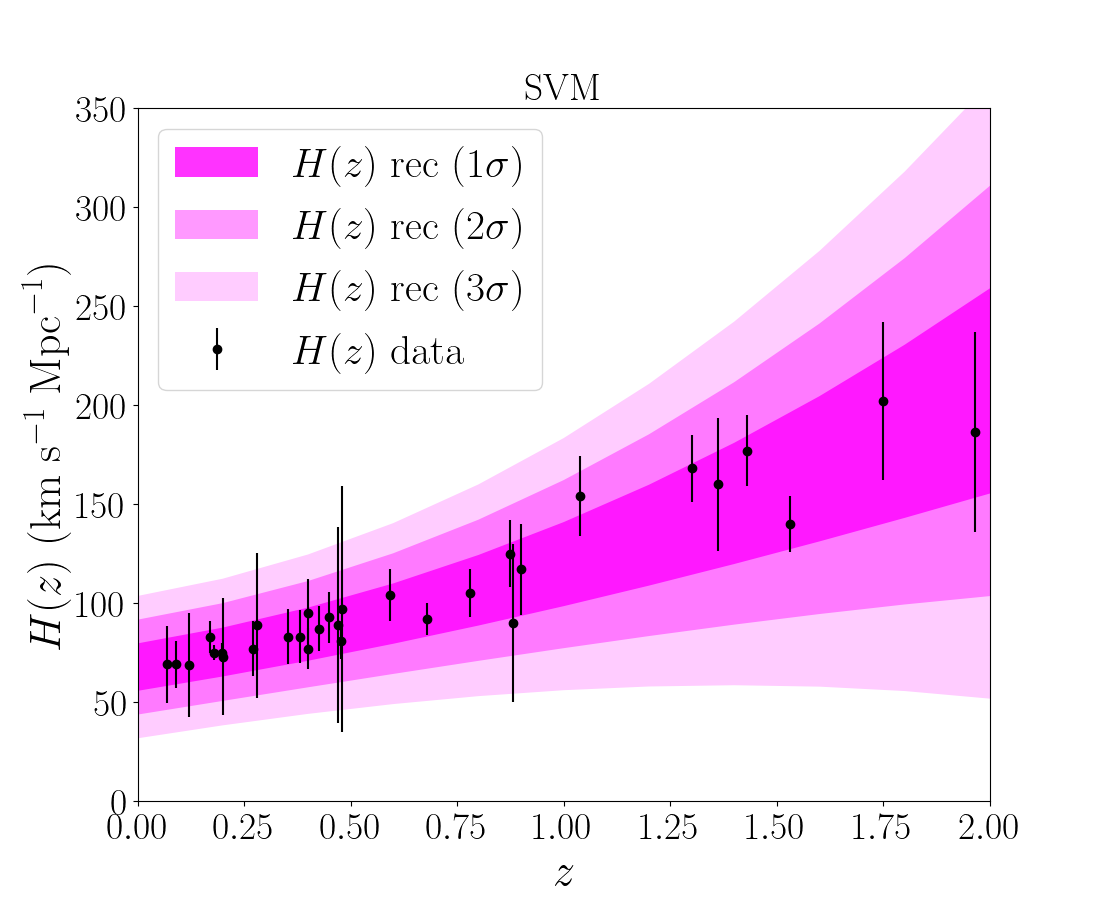}
\includegraphics[width=0.49\textwidth, height=7.3cm]{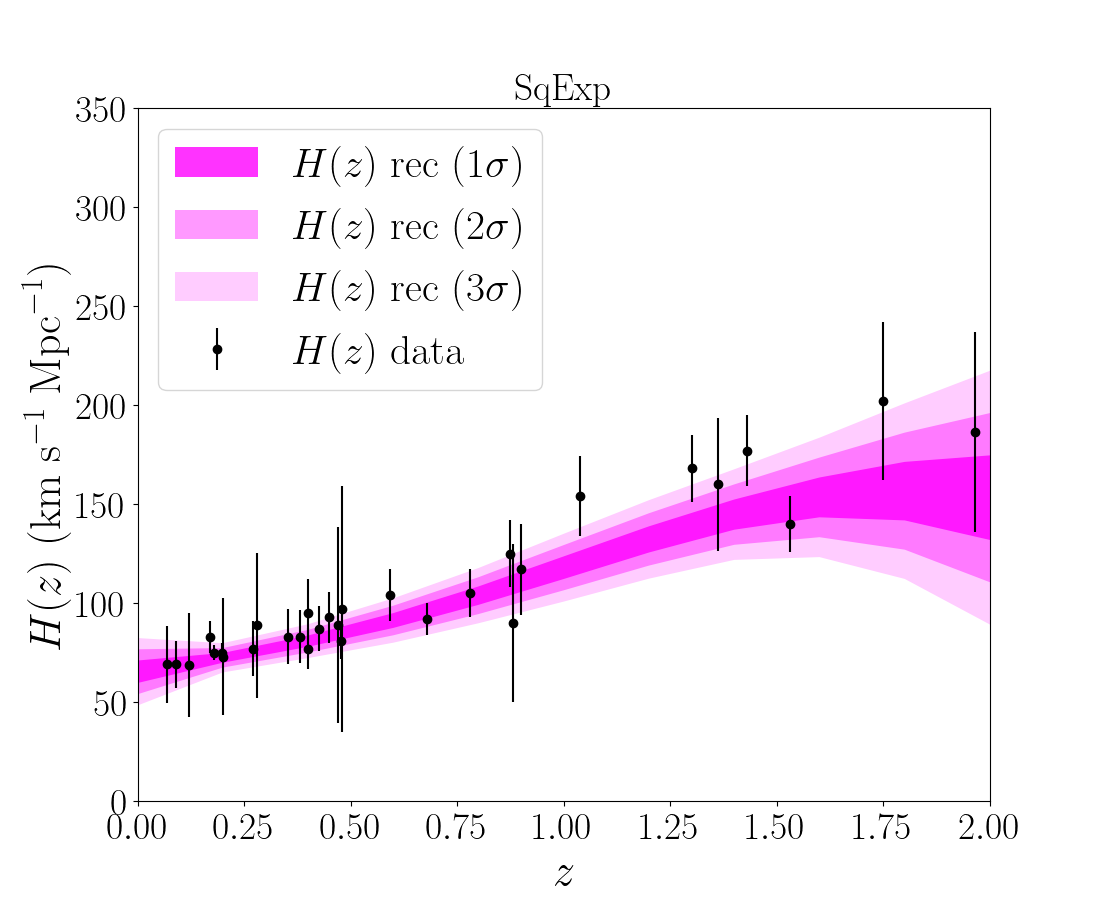}
\includegraphics[width=0.49\textwidth, height=7.3cm]{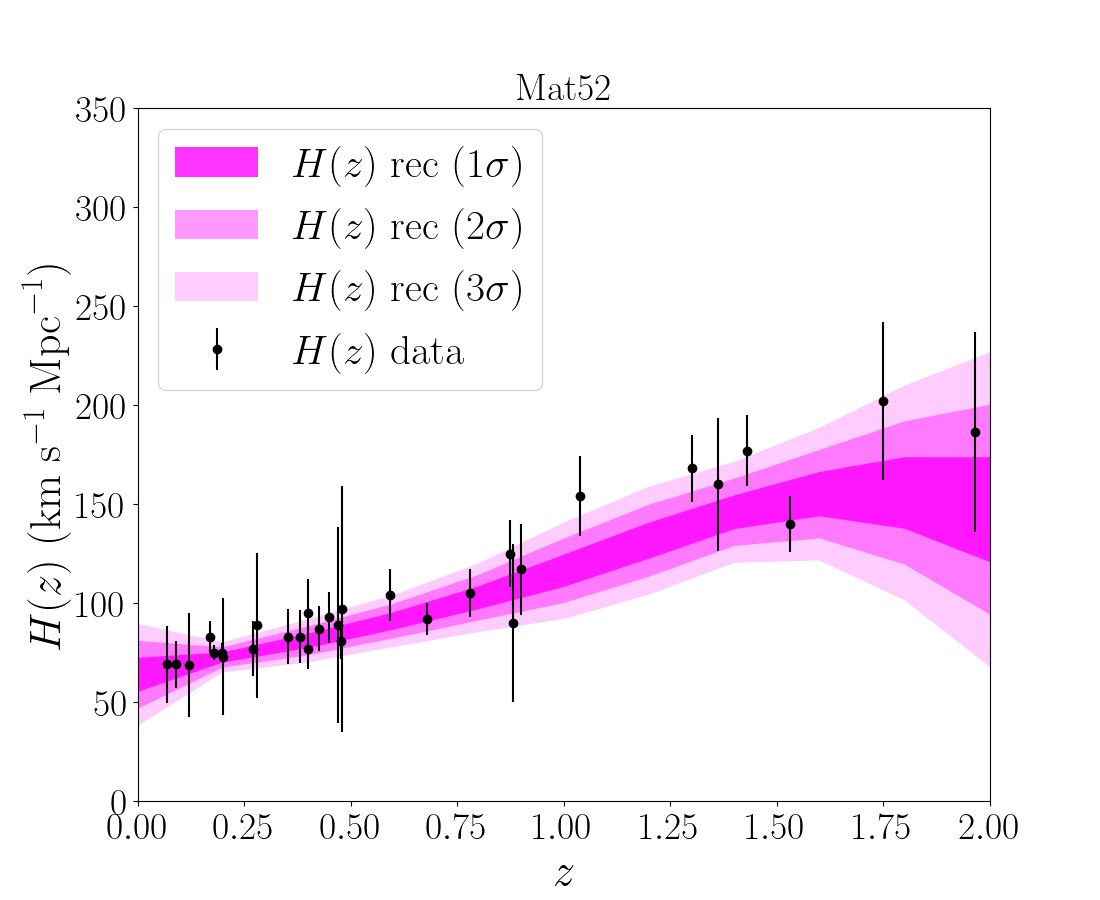}
\caption{The reconstructed $H(z)$ values obtained from EXT (top left), ANN (top right), GBR (center left), SVM (center right), SqExp (lower left) and Mat52 (lower right). Different shades of magenta represent different confidence level for the reconstructions, ranging from $1$ (darker shade) to $3\sigma$ (lighter shade).}
\label{fig:hz_reconst}
\end{figure*}


We show our $H_0$ measurements for each algorithm in Fig.~\ref{fig:H0_measurements}. The top panels present the results obtained from the EXT (left) and ANN (right) algorithms, whereas the middle panels display the GBR (left) and SVM (right) results, and the bottom ones show the GAP predictions for the Mat52 and SqExp kernels in the left and right plots, respectively. Each data point at these plots represents the predicted $H_0$ values ($H^{\rm pred}_0$) according to the prescription described in Section IIB for each simulated data-set specifications, i.e., different $\sigma_{H}/H$ results against the $N_z$ values. The light blue horizontal line corresponds to the fiducial $H_0$. We can see that GBR and EXT are able to correctly predict the fiducial $H_0$ for higher $N_z$ and lower $\sigma_{H}/H$ values, but not otherwise - especially for low values of $N_z$, where these algorithms predict a larger $H_0$ value. This indicates a bias in the results, despite the cross-validation procedure adopted. We also find that ANN and SVM are able to recover the fiducial $H_0$, even for lower quality sets of simulations. However, its predictions present larger variances as $\sigma_{H}/H$ increases.

These results show that GBR and EXT are more sensitive to $N_z$ concerning bias, whereas ANN is more sensitive to $\sigma_{H}/H$ with respect to its variance. On the other hand, SVM exhibits the best bias-variance tradeoff among all algorithms, as shown in Fig.~\ref{fig:BVT_measurements}, along with tables~\ref{tab:results_sigh0p008} to~\ref{tab:results_sigh0p08}, presented in Appendix B. We obtain that SVM is able to recover the fiducial $H_0$ without significant losses in bias and variance as the simulation quality decreases. Such a result may happen due to a few reasons: For example, due to the non-guaranteed convergence of neural networks. By an appropriate choice of hyperparameters, ANN can approach a target function until a satisfactory result is reached; however, SVMs are theoretically grounded in their capacity to converge to the solution for a problem. Note that we adopted a polynomial kernel for SVM, and a nonlinear activaction function for ANN, namely "relu", in order to make a fair comparison between them. As for the EXT case, such an algorithm is known to be prone to overfitting and to outliers, which can explain the larger bias with lower variance. A similar problem applies for the GBR case as well. Not to mention that both demand a longer training time than ANN and SVM, which translates into a longer computational time to obtain the $H_0$ measurements and uncertainties in our case. 

Regarding the comparison with the results obtained with GAP using the Squared Exponential (SqExp) and Mat\'ern(5/2) (Mat52) kernels, we find good agreement between the SVM and GAP measurement of $H_0$, in spite of a slightly larger BVT for the former. But note that GAP can also be prone to overfitting, as the data points with smaller uncertainties have a greater impact to determine the function that best represents the distance between data points to perform its numerical reconstruction - especially when assuming the SqExp kernel, which presents greater differentiability than the Mat52 case. Therefore, we show that SVM can be used as a cross-check method for GAP regression, which has been widely used in the literature.. 

Moreover, we show the $H(z)$ reconstructions obtained from the simulations mimicking the real data configuration in Fig.~\ref{fig:hz_reconst}, at a $1$, $2$ and $3\sigma$ confidence level, alongside the actual $H(z)$ measurements. We can clearly see a "step-wise" behaviour on the EXT and GBR reconstructions, contrarily to other algorithms. This illustrates the bias problems they face, as commented before. Once again, the GAP results exhibit the smallest uncertainties among all, but this may also happen due to possible overfitting, as commented before. This is exemplified by the dip on the high-$z$ end of the reconstruction. Nevertheless, the Hubble Constant measured by all algorithms are in agreement with each other, as depicted in Table~\ref{tab:results_sighz}, where EXT and GBR again exhibit a tendency towards larger $H_0$ values - and hence larger BVT - while ANN and SVM present less biased results, as in the previous cases. Interestingly, we find that ANN performed slightly better than SVM this time around, yielding a slightly lower BVT. Note also that our results are in good agreement with the predicted $H_0$ in~\cite{wang20a}, who used ANN as well in their analysis\footnote{The authors reported $H_0 = 67.33 \pm 15.74 \; \mathrm{km \; s}^{-1} \; \mathrm{Mpc}^{-1}$.}, but we could obtain a slightly lower uncertainty in our predictions - roughly 17\% uncertainty versus 23\% in their case.

We also checked whether the test sample size affects the $H_0$ predictions and their bias-variance tradeoff. We find that its default choice, i.e., a split between 75\%-25\% between training and test sample, respectively, provides the best results for all algorithms compared to a 90\%-10\% and 60\%-40\% split, for instance. The EXT and GBR algorithms perform similarly for 10\% and 25\%, but their predictions become significantly worse for the 40\% case. This is an expected result, since both algorithms require large training sets to carry out such predictions. On the other hand, ANN and SVM perform significantly worse for split choices other than the default one. Finally, we verified the results for different cross-validations values, such as CV$=2,4,8$, finding consistent values with those obtained with the standard choice CV$=3$. 

\section{Conclusions}

Machine learning has been gaining formidable importance in present times. Given the state-of-art of modern computation and processing facilities, the application of machine learning algorithms in physical sciences not only became popular, but essential in the process of handling huge data-sets and performing model prediction - specially in light of forthcoming redshift surveys.

Our work focused on a comparison of different machine learning algorithms for the sake of measuring the Hubble Constant from cosmic chronometers measurements. We used four different algorithms in our analysis, which are based on decision trees, artificial neural networks, support vector machine and gradient boosting, as available in the \texttt{scikit-learn} python package. We applied them on simulated $H(z)$ data-sets assuming different specifications, and assuming a flat $\Lambda$CDM model consistent with Planck 2018 best fit, in order to measure $H_0$ through an extrapolation procedure. 

Our uncertainties are estimated using a Monte Carlo-bootstrap method on the simulations, after properly splitting them into training and test sets, and performed a grid search over their hyperparameter space during the cross-validation procedure. In addition, we created a performance ranking between these methods via the bias-variance tradeoff, and compared them with other established methods in the literature, e.g. Gaussian Processes as in the {\sc GaPP} code. 

We obtained that the algorithms based on decision trees and gradient boosting present the lowest performance among all, as they provide low variance with a large bias in the reconstructed $H_0$. Instead, the artificial neural networks and support vector machine are able to correctly recover the fiducial $H_0$ value, where the latter method exhibits the lowest variance among them. We also found that the support vector machine algorithm presents compatible benchmark metrics with the Gaussian Processes one. This result shows that such method can be successfully used as a cross-check method between different non-parametric reconstruction techniques, which will be of great importance in the advent of next-generation cosmological surveys~\cite{jpas14,minijpas21a,desi16,euclid18,ska20,lsst18}, as they are expected to provide $H(z)$ measurements with a few percent precision.

\section*{Acknowledgements}
We thank the annonymous referee for constructive criticism on the manuscript. CB acknowledges financial support from the FAPERJ postdoc nota 10 fellowship. LC acknowledges financial support from CNPq (Grant No. 310314/2019-4). J. Alcaniz is supported by Conselho Nacional de Desenvolvimento Cient\'{\i}fico e Tecnol\'ogico CNPq (Grants no. 310790/2014-0 and 400471/2014-0) and Funda\c{c}\~ao de Amparo \`a Pesquisa do Estado do Rio de Janeiro FAPERJ (grant no. 233906). We thank the National Observatory Data Center (CPDON) for computational support.

\section*{Data Availability Statement}
The associated data and scripts developed in this work can be found at: \url{https://github.com/astrobengaly/machine_learning_H0_v2}.


\begin{appendix}

\section{Algorithm performance results}

\begin{table*}[!h]
\begin{tabular}{cccccc}
\hline 
\hline 
\; alg \; & \;\; $N_z$ \;\; & $H_0 \pm \sigma_{H_0}$ \quad & BVT\\
\hline
\hline 
    & $20$ & $69.443 \pm 2.151$ \; & $8.966$ \\
EXT & $30$ & $68.593 \pm 1.385$ \; & $3.438$  \\
    & $50$ & $68.288 \pm 0.877$ \; & $1.632$  \\
    & $80$ & $67.844 \pm 0.607$ \; & $0.601$ \\ 
\hline
     & $20$ & $67.111 \pm 1.695$ \; & $2.936$ \\
ANN & $30$ & $67.179 \pm 1.071$ \; & $1.180$ \\
     & $50$ & $67.198 \pm 0.606$ \; & $0.395$ \\
     & $80$ & $67.262 \pm 0.456$ \; & $0.217$ \\
\hline
    & $20$ & $70.441 \pm 3.456$ \; & $21.438$ \\
GBR & $30$ & $69.837 \pm 2.857$ \; & $14.296$ \\
    & $50$ & $68.747 \pm 1.673$ \; & $4.723$ \\
    & $80$ & $68.170 \pm 1.124$ \; & $1.920$ \\
\hline
    & $20$ & $66.973 \pm 0.892$ \; & $0.946$ \\
SVM & $30$ & $67.004 \pm 0.698$ \; & $0.614$ \\
    & $50$ & $67.090 \pm 0.533$ \; & $0.357$ \\
    & $80$ & $67.181 \pm 0.468$ \; & $0.251$ \\ 
\hline
\hline
	& $20$ & $67.261 \pm 0.527$ \; & $0.287$ \\
SqExp & $30$ & $67.286 \pm 0.428$ \; & $0.188$ \\
    & $50$ & $67.374 \pm 0.290$ \; & $0.084$ \\
	  & $80$ & $67.343 \pm 0.215$ \; & $0.046$ \\
\hline
	  & $20$ & $67.189 \pm 0.792$ \; & $0.657$ \\
Mat52 & $30$ & $67.312 \pm 0.545$ \; & $0.302$ \\
	  & $50$ & $67.351 \pm 0.486$ \; & $0.236$ \\
	  & $80$ & $67.363 \pm 0.321$ \; & $0.103$ \\ 
\hline
\hline
\end{tabular}
\caption{Respectively: $H_0$ measurements (in units of $\mathrm{km s}^{-1} \; \mathrm{Mpc}^{-1}$), as well as the BVT values obtained for all $N_z$ and ML algorithms assuming $\sigma_H/H=0.008$. We also show the results obtained using the GAP method at the bottom of the table.} 
\label{tab:results_sigh0p008} 
\end{table*}

\begin{table*}[!h]
\begin{tabular}{cccccc}
\hline 
\hline 
\; alg \; & \;\; $N_z$ \;\; & $H_0 \pm \sigma_{H_0}$ \quad & BVT\\
\hline
\hline 
    & $20$ & $69.678 \pm 2.334$ \; & $10.818$ \\
EXT & $30$ & $68.491 \pm 1.132$ \; & $2.560$ \\
    & $50$ & $68.157 \pm 1.111$ \; & $1.869$ \\
    & $80$ & $67.928 \pm 0.796$ \; & $0.955$ \\ 
\hline
     & $20$ & $66.999 \pm 1.224$ \; & $1.629$ \\
ANN & $30$ & $67.064 \pm 0.743$ \; & $0.639$ \\
     & $50$ & $67.148 \pm 0.767$ \; & $0.637$ \\
     & $80$ & $67.255 \pm 0.579$ \; & $0.346$ \\
\hline
    & $20$ & $70.682 \pm 3.140$ \; & $20.894$ \\
GBR & $30$ & $69.276 \pm 2.009$ \; & $7.706$ \\
    & $50$ & $68.431 \pm 1.398$ \; & $3.103$ \\
    & $80$ & $68.325 \pm 1.324$ \; & $2.685$ \\
\hline
    & $20$ & $67.005 \pm 0.930$ \; & $0.990$ \\
SVM & $30$ & $66.990 \pm 0.812$ \; & $0.796$ \\
    & $50$ & $67.074 \pm 0.696$ \; & $0.566$ \\
    & $80$ & $67.179 \pm 0.537$ \; & $0.321$ \\ 
\hline
\hline
	& $20$ & $67.321 \pm 0.708$ \; & $0.503$ \\
SqExp & $30$ & $67.286 \pm 0.428$ \; & $0.188$ \\
    & $50$ & $67.300 \pm 0.352$ \; & $0.127$ \\
	  & $80$ & $67.354 \pm 0.282$ \; & $0.080$ \\
\hline
	  & $20$ & $67.011 \pm 1.270$ \; & $1.735$ \\
Mat52 & $30$ & $67.213 \pm 0.627$ \; & $0.415$ \\
	  & $50$ & $67.183 \pm 0.488$ \; & $0.270$ \\
	  & $80$ & $67.352 \pm 0.442$ \; & $0.195$ \\
\hline
\hline
\end{tabular}
\caption{Same as Table~\ref{tab:results_sigh0p008}, but assuming $\sigma_H/H=0.01$.}\label{tab:results_sigh0p01} 
\end{table*}

\begin{table*}[!h]
\begin{tabular}{cccccc}
\hline 
\hline 
\; alg \; & \;\; $N_z$ \;\; & $H_0 \pm \sigma_{H_0}$ \quad & BVT\\
\hline
\hline 
    & $20$ & $69.787 \pm 2.461$ \; & $11.946$ \\
EXT & $30$ & $69.407 \pm 2.120$ \; & $8.683$ \\
    & $50$ & $68.762 \pm 2.007$ \; & $5.995$ \\
    & $80$ & $68.377 \pm 1.627$ \; & $3.682$ \\ 
\hline
     & $20$ & $67.140 \pm 2.997$ \; & $9.031$ \\
ANN & $30$ & $67.610 \pm 2.528$ \; & $6.451$ \\
     & $50$ & $67.059 \pm 2.001$ \; & $4.094$ \\
     & $80$ & $67.169 \pm 1.321$ \; & $1.782$ \\
\hline
    & $20$ & $70.906 \pm 3.589$ \; & $25.457$ \\
GBR & $30$ & $70.566 \pm 2.967$ \; & $19.083$ \\
    & $50$ & $69.236 \pm 2.145$ \; & $8.118$ \\
    & $80$ & $68.843 \pm 1.847$ \; & $5.611$ \\
\hline
    & $20$ & $67.016 \pm 2.332$ \; & $5.557$ \\
SVM & $30$ & $67.152 \pm 1.843$ \; & $3.440$ \\
    & $50$ & $66.538 \pm 1.443$ \; & $2.758$ \\
    & $80$ & $66.879 \pm 1.153$ \; & $1.561$ \\ 
\hline
\hline
	  & $20$ & $66.690 \pm 1.574$ \; & $2.925$ \\
SqExp & $30$ & $67.063 \pm 1.238$ \; & $1.622$ \\
    & $50$ & $67.021 \pm 0.902$ \; & $0.929$ \\
	  & $80$ & $67.177 \pm 0.753$ \; & $0.601$ \\
\hline
	  & $20$ & $66.963 \pm 1.998$ \; & $4.152$ \\
Mat52 & $30$ & $67.313 \pm 1.657$ \; & $2.746$ \\
	& $50$ & $67.228 \pm 1.270$ \; & $1.631$ \\
	& $80$ & $67.224 \pm 1.049$ \; & $1.118$ \\ 
\hline
\hline
\end{tabular}
\caption{Same as Table~\ref{tab:results_sigh0p008}, but assuming $\sigma_H/H=0.03$.} \label{tab:results_sigh0p03} 
\end{table*}

\begin{table*}[!h]
\begin{tabular}{cccccc}
\hline 
\hline 
\; alg \; & \;\; $N_z$ \;\; & $H_0 \pm \sigma_{H_0}$ \quad & BVT\\
\hline
\hline 
    & $20$ & $70.189 \pm 3.672$ \; & $21.486$ \\
EXT & $30$ & $69.039 \pm 3.485$ \; & $14.964$ \\
    & $50$ & $68.805 \pm 2.601$ \; & $8.850$ \\
    & $80$ & $68.281 \pm 2.596$ \; & $7.586$ \\ 
\hline
    & $20$ & $67.445 \pm 4.754$ \; & $22.606    $ \\
ANN & $30$ & $66.417 \pm 3.212$ \; & $11.207$ \\
    & $50$ & $67.272 \pm 2.361$ \; & $5.580$ \\
    & $80$ & $67.460 \pm 1.818$ \; & $3.316$ \\
\hline
    & $20$ & $71.122 \pm 4.001$ \; & $30.164$ \\
GBR & $30$ & $69.474 \pm 3.415$ \; & $16.128$ \\
    & $50$ & $69.374 \pm 2.864$ \; & $12.258$ \\
    & $80$ & $68.938 \pm 2.847$ \; & $10.596$ \\
\hline
    & $20$ & $66.111 \pm 3.831$ \; & $16.238$ \\
SVM & $30$ & $65.993 \pm 2.894$ \; & $10.242$ \\
    & $50$ & $66.680 \pm 1.772$ \; & $3.602$ \\
    & $80$ & $66.881 \pm 1.756$ \; & $3.311$ \\ 
\hline
\hline
	& $20$ & $66.553 \pm 2.252$ \; & $5.722$ \\
SqExp & $30$ & $66.153 \pm 1.784$ \; & $4.641$ \\
	  & $50$ & $66.820 \pm 1.338$ \; & $2.082$ \\
	  & $80$ & $66.900 \pm 1.005$ \; & $1.221$ \\
\hline
	  & $20$ & $66.920 \pm 2.914$ \; & $8.684$ \\
Mat52 & $30$ & $66.442 \pm 2.340$ \; & $6.318$ \\
	  & $50$ & $66.918 \pm 1.753$ \; & $3.269$ \\
	  & $80$ & $67.141 \pm 1.350$ \; & $1.872$ \\ 
\hline
\hline
\end{tabular}
\caption{Same as Table~\ref{tab:results_sigh0p008}, but assuming $\sigma_H/H=0.05$.} 
\label{tab:results_sigh0p05} 
\end{table*}

\begin{table*}[!h]
\begin{tabular}{cccccc}
\hline 
\hline 
\; alg \; & \;\; $N_z$ \;\; & $H_0 \pm \sigma_{H_0}$ \quad & BVT\\
\hline
\hline 
    & $20$ & $71.322 \pm 4.424$ \; & $35.271$ \\
 EXT & $30$ & $70.248 \pm 5.092$ \; & $34.274$ \\
    & $50$ & $70.305 \pm 3.586$ \; & $21.535$ \\
    & $80$ & $69.040 \pm 4.267$ \; & $21.028$ \\ 
\hline
    & $20$ & $68.874 \pm 8.605$ \; & $76.333$ \\
 ANN & $30$ & $66.293 \pm 7.910$ \; & $63.698$ \\
    & $50$ & $67.952 \pm 5.531$ \; & $30.943$ \\
    & $80$ & $67.143 \pm 3.521$ \; & $12.447$ \\
\hline
    & $20$ & $72.979 \pm 4.907$ \; & $55.656$ \\
 GBR & $30$ & $72.081 \pm 5.080$ \; & $48.096$ \\
    & $50$ & $71.026 \pm 4.250$ \; & $31.501$ \\
    & $80$ & $69.742 \pm 3.948$ \; & $21.261$ \\
\hline
    & $20$ & $66.813 \pm 4.799$ \; & $23.333$ \\
SVM & $30$ & $66.271 \pm 5.076$ \; & $26.949$ \\
    & $50$ & $66.395 \pm 3.578$ \; & $13.731$ \\
    & $80$ & $66.563 \pm 2.987$ \; & $9.557$ \\ 
\hline
\hline
    & $20$ & $66.014 \pm 3.378$ \; & $13.224$ \\
SqExp & $30$ & $66.130 \pm 2.652$ \; & $8.544$ \\
	& $50$ & $66.611 \pm 1.963$ \; & $4.413$ \\
	  & $80$ & $65.910 \pm 1.829$ \; & $5.448$ \\
\hline
	  & $20$ & $66.234 \pm 3.832$ \; & $15.955$ \\
Mat52 & $30$ & $66.644 \pm 3.147$ \; & $10.418$ \\
	& $50$ & $67.085 \pm 2.471$ \; & $6.183$ \\
	  & $80$ & $66.369 \pm 2.316$ \; & $6.344$ \\ 
\hline
\hline
\end{tabular}
\caption{Same as Table~\ref{tab:results_sigh0p008}, but assuming $\sigma_H/H=0.08$.}
\label{tab:results_sigh0p08} 
\end{table*}

\begin{table*}[]
\begin{tabular}{cccccc}
\hline 
\hline 
\; alg \; & $H_0 \pm \sigma_{H_0}$ \quad & BVT\\
\hline
\hline 
EXT & $70.829 \pm 12.839$ \; & $176.881$ \\
ANN & $68.412 \pm 11.389$ \; & $130.821$ \\
GBR & $71.766 \pm 14.255$ \; & $222.620$ \\
SVM & $67.601 \pm 11.982$ \; & $143.634$ \\ 
\hline
\hline
SqExp & $66.890 \pm 6.065$ \; & $37.005$ \\
Mat52 & $66.355 \pm 9.568$ \; & $92.556$ \\
\hline
\hline
\end{tabular}
\caption{Same as Table~\ref{tab:results_sigh0p008}, but assuming the real data configuration, i.e., $N_z=31$ and the same $\sigma_{H(z)}$ of the real $H(z)$ measurements.}
\label{tab:results_sighz} 
\end{table*}

\end{appendix}

\end{document}